\documentclass[%
nofootinbib,
 amsmath,amssymb,
 aps,
prd,
twocolumn
]{revtex4-2}

\usepackage{graphicx}
\usepackage{dcolumn}
\usepackage{bm}
\usepackage{xcolor}

\begin{document}

\title{Stability of hybrid stars via catastrophe theory}

\author{Eduardo S. Fraga}
 \email{fraga@if.ufrj.br}

\author{Sergio E. Jor\'as}%
 \email{joras@if.ufrj.br}
 
\affiliation{%
  Instituto de F\'\i sica, Universidade Federal do Rio de Janeiro,\\
 CEP 21941-972 Rio de Janeiro, RJ, Brazil}
%


\begin{abstract}

We discuss the possible topologies of the mass-radius relation for hybrid stars in light of catastrophe theory. We show that different branches of compact stars emerge or disappear depending on the bifurcation structure of the underlying catastrophe potential function, the mathematical analog of an effective potential. For that we assume a single first-order phase transition between nuclear and quark matter. After a warm up with incompressible fluids, we present a qualitative discussion explaining the (dis)appearance of (un)stable branches for hybrid stars as the control parameters change.


\end{abstract}

\maketitle


\section{Introduction}
\label{sec:intro}

Cold neutron star matter can, in principle, encompass different phases of strong interactions \cite{glendenning1997compact,Schaffner-Bielich:2020psc}. The extent to which one would find a core of cold quark matter inside a hadronic mantle, forming a hybrid star, depends essentially on the full equation of state provided by Quantum Chromodynamics (QCD). However, a nonperturbative first-principle solution at nonzero baryon density seems still far away, given the well-known Sign Problem of lattice QCD \cite{Hands:2007by,deforcrand2009pos}. Nevertheless, as we will discuss in the following, the stability of hybrid stars might be less sensitive to details of the underlying equations of state, and more to structural properties such as their singularity profile, especially in the case of a sharp first-order phase transition.

Equilibrium compact star configurations are given by the solutions of the Tolman-Oppenheimer-Volkov (TOV) equations cite{Tolman:1939jz,Oppenheimer:1939ne}, which assume the validity of General Relativity (GR), spherical symmetry, and hydrostatic equilibrium \cite{glendenning1997compact,Schaffner-Bielich:2020psc}. The families of possible solutions are obtained once an equation of state, which encodes the microscopic information, is provided. For each central value of the pressure or energy density, there exists a unique non-singular solution for a given equation of state. However, even if those solutions represent  equilibrium states of the stellar configuration, they may be states of stable or unstable configurations. In order to tell if a given configuration is stable, one would generally have to study their normal modes \cite{Weinberg:1972kfs}. Nevertheless,
one can usually consider a given stellar configuration stable as long as it satisfies the static stability criterion
$\partial{M}/\partial{\epsilon_c}\geq 0$, where $\epsilon_c$ is the central energy density, even though the original theorem applies strictly to a perfect fluid with constant chemical composition and entropy per nucleon \cite{Weinberg:1972kfs,harrison1965gravitation}.

The stability of hybrid stars in the presence of a first-order phase transition has been addressed previously in different contexts \cite{Seidov1967,seidov1971stability,Kampfer:1981zmq,schaeffer1983phase,Zdunik1987,lindblom1998phase,Alford:2013aca}. In particular, Ref. \cite{lindblom1998phase} investigates how the properties of the phase transition might be extracted from the structure of the mass-radius curve. Generic conditions for the structure and stability of hybrid stars have been proposed in Ref. \cite{Alford:2013aca}. Using a parametrization of quark matter based on a constant speed of sound, the authors explore how quantities such as the transition pressure $p_{\rm tr}$, the energy density discontinuity $\Delta\epsilon$, and the stiffness of quark matter affect the resulting mass–radius relation. This allows them to identify the conditions under which stable hybrid star branches can appear, including the possibility of disconnected “twin star” solutions. They also demonstrate that hybrid stars compatible with observed $2M_\odot$ pulsars can be realized within a broad region of their parameter space. See also Ref. \cite{Christian:2017jni} for a classification of twin star solutions for a constant speed of sound parametrized equation of state.

In this paper we follow a different path that seems to be overlooked in previous analyzes. We discuss the possible topologies of the mass-radius relation for hybrid stars in light of {\it catastrophe theory}  \cite{saunders1980catastrophe}, which classifies and studies how the extrema of certain functions coalesce and
emerge. We show that different branches of compact stars emerge or disappear depending on the bifurcation structure of the underlying catastrophe potential function, the mathematical analog of an effective potential. This means that the structure of such singularities is what ultimately defines the topology of the mass-radius diagram. To implement this analysis we assume a single first-order phase transition between nuclear and quark matter. We take it to be a sharp transition, so that there is no need for a Gibbs construction. Given most of the current results for the surface tension in neutron star matter \cite{Palhares:2010be,Pinto:2012aq,Mintz:2012mz,Lugones:2013ema,Fraga:2018cvr,Schmitt:2020tac,Fraga:2023wtd}, this seems to be a reasonable working hypothesis.

The paper is organized as follows. For completeness, in Section \ref{sec:planets} we discuss the instability of planetary cores and of stars with a phase transition. The latter corresponds at the end to the Seidov criterion. Then, we introduce the idea of bi-homogeneous hybrid stars in Section \ref{sec:bi-homogeneous}, a warm up system with incompressible fluids, where a great deal of the calculations can be performed analytically to illustrate the basic points. In Section \ref{sec:catastrophe} we introduce the basic elements of catastrophe theory that will underlie our framework. Finally, we present the final discussion in Section \ref{sec:dots}, within the framework of catastrophe theory without specifying the equation of state for each phase. We close with our summary and outlook in Section \ref{sec:summary}.

\section{From the instability of small planetary cores to phase transitions in hybrid stars}
\label{sec:planets}

\subsection{TOV equations}
\label{sec:TOV}

The structure of a compact star is given by the solution of the Tolman-Oppenheimer-Volkov (TOV) equations \cite{Tolman:1939jz,Oppenheimer:1939ne,glendenning1997compact,Schaffner-Bielich:2020psc}, which encode Einstein’s General Relativity (GR) field equations in hydrostatic equilibrium for a spherical geometry. In natural units, such that $G=1$ and $c=1$, we have
\begin{align}
\frac{dp}{dr} &= -\frac{m}{r^2}\big(\epsilon + p \big)
\left( 1 + \frac{4\pi r^3 p}{m}\right)\left(1-\frac{2m}{r} \right)^{-1} \; , \\
\frac{dm}{dr} &= 4 \pi r^2 \epsilon \; ,
\end{align}
where the explicit radial dependence of the functions $m(r)$, $\epsilon(r)$, $p(r)$ --- mass, energy density and pressure, respectively ---  has been dropped to avoid over cluttering.  

Given the equation of state $p = p(\epsilon)$, one can integrate the TOV equations from the origin ($r=0$) until the pressure vanishes, $p(r=R) = 0$, where $R$ is the radius of the star, starting from the boundary conditions at the center:
\begin{align}
p(r=0) &=p_c\\
\frac{dp}{dr}\Big|_{r=0} &= 0 \; .
\end{align}
Of course, one can also choose to use different functions evaluated at $r=0$, such as the central energy density or the value of the baryon chemical potential at the center of the star, to perform the integration.

Different equations of state will produce different types of stars (white dwarfs,
neutron stars, strange stars, quark stars, etc), and curves on the mass-radius diagram for the families of stars \cite{glendenning1997compact,Schaffner-Bielich:2020psc}.

%
%

\begin{figure}[t]
\includegraphics[width=0.4\textwidth]{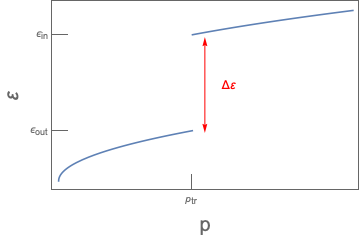}
\caption{Qualitative behavior of the energy density as a function of the pressure for a hybrid star. Note the gap $\Delta \epsilon\equiv \epsilon_{\rm in}-\epsilon_{\rm out}$ at $p=p_{\rm tr}$, at the boundary $r=r_{\rm tr}$ between the two phases. }
\label{gap}
\end{figure}

In the case of hybrid stars in the presence of a first-order phase transition that happens at a pressure $p=p_{\rm tr}$, there will be a boundary between the two layers of the star, at $r_{\rm tr}$. Then, the boundary conditions on this surface are given by (see Fig.~\ref{gap})
\begin{align}
\epsilon(r_{\rm tr}^-) &= \epsilon(r_{\rm tr}^+) + \Delta\epsilon  \; , \\
p(r_{\rm tr}^-) &= p(r_{\rm tr}^+) = p_{\rm tr} \;,
\end{align}
where $\Delta\epsilon>0$ is the energy gap at the phase transition -- see Fig.~\ref{gap}.

\subsection{The Seidov criterion}
\label{sec:seidov}

The fact that a first-order phase transition in neutron star matter can drive an entire region of the mass-radius diagram unstable was shown over fifty years ago by Seidov \cite{Seidov1967,seidov1971stability}. Instabilities in the Newtonian case, which are relevant for small planetary cores, were previously investigated in Refs. \cite{Ramsey1950,Lighthill1950}. In the case of neutron stars, the presence of an instability that might generate new possible solutions (a new stable branch) in a bifurcation process is a necessary condition for the appearance of a so-called third family of compact stars \cite{Schaffner-Bielich:2020psc}.

To illustrate how the Seidov criterion emerges, let us consider the simpler case of a hybrid star made of two homogeneous incompressible fluids with energy densities $\epsilon_{\rm out}$ and $\epsilon_{\rm in}$, where $\epsilon_{\rm in} > \epsilon_{\rm out}$. That would correspond to the situation displayed in Fig. \ref{gap}, but with two flat curves for $\epsilon(p)$. This condition can be easily relaxed and serves only the purpose of simplifying the notation here. The results presented below are general.

If we write the central pressure as
\begin{equation}
    p_c=p_{\rm tr} + \delta \; ,
\end{equation}
and impose the boundary conditions
\begin{equation}
    p(r_{\rm tr})=p_{\rm tr} \;,
\end{equation}
\begin{equation}
    m(r_{\rm tr})=\frac{4}{3}\pi r_b^3 \epsilon_{\rm in} \;,
\end{equation}
we can solve the TOV equations and obtain $r_b$ as a function of $\delta$ for small $\delta$ ($\delta \ll p_{\rm tr}$)
\begin{equation}
    r_b^2 \approx \frac{3}{2\pi}\frac{\delta}{(\epsilon_{\rm in} + 3p_{\rm tr})(\epsilon_{\rm in} + p_{\rm tr})} \;,
\end{equation}
which recovers the result originally obtained by Seidov \cite{seidov1971stability}.

Approaching the case in the presence of a phase transition as a perturbation of the case in its absence, one can compute the perturbation function $\Pi (r) \equiv p^+(r) - p^-(r)$, where ($+/-$) corresponds to the case with/without a phase transition. For small values of $r$, we have the form \cite{seidov1971stability}
\begin{equation}
    \Pi(r)=A + \frac{B}{r} \;,
\end{equation}
with 
\begin{equation}
    A= \left[ \frac{(\epsilon_{\rm out}+p_{\rm tr})(3\epsilon_{\rm out}-2\epsilon_{\rm in}+3p_{\rm tr})}{(\epsilon_{\rm in}+p_{\rm tr})(\epsilon_{\rm in}+3p_{\rm tr})}  \right] \delta
\end{equation}
being the only relevant coefficient for the behavior of the total mass function, since $B/r$ vanishes in regions far from the (small) nucleus.

The right and left derivatives of the mass with respect to the central pressure $p_c=p_{\rm tr}$ are related by
\begin{equation}
    \left( \frac{dM}{dp_c}\right)^{+}= \left[ \frac{(\epsilon_{\rm out}+p_{\rm tr})(3\epsilon_{\rm out}-2\epsilon_{\rm in}+3p_{\rm tr})}{(\epsilon_{\rm in}+p_{\rm tr})(\epsilon_{\rm in}+3p_{\rm tr})}  \right] \left( \frac{dM}{dp_c}\right)^{-} \;,
\end{equation}
so that the derivative of $M$ with respect to $p_c$ changes sign, signaling the appearance of an instability, for 
\begin{equation}
    \frac{\epsilon_{\rm in}}{\epsilon_{\rm out}}>\frac{3}{2}\left( 1+ \frac{p_{\rm tr}}{\epsilon_{\rm out}}  \right)\;,
\end{equation}
which corresponds to the Seidov criterion \cite{seidov1971stability}.

\section{Bi-homogeneous hybrid stars}
\label{sec:bi-homogeneous}

As we mentioned previously, the possible topologies of the mass-radius relation for hybrid stars, i.e., the fact that different branches of stars can emerge or disappear, depends on the bifurcation structure of an underlying potential function. In other words, its structure of singularities is what ultimately defines the topology of the mass-radius diagram. The natural framework to investigate the set of singularities is provided by {\it catastrophe theory}.

\begin{figure}
\includegraphics[width=0.4\textwidth]{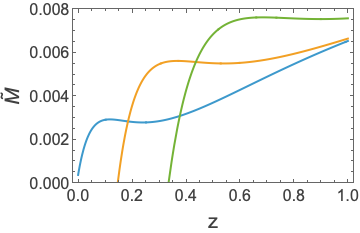}
\includegraphics[width=0.4\textwidth]{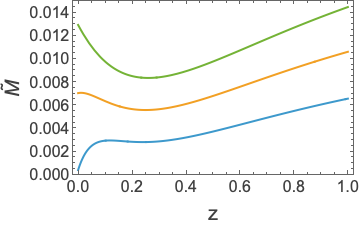}
\caption{Curves $\tilde M \times z\equiv p_c/\epsilon_{\rm out}$ for 
(top) $x=0.1, 0.2, 0.3$ (from left to right) and $y=0.1$ and (bottom) $x=0.1$ and $y=0.1, 0.2, 0.3$ (from bottom to top). We recall the reader that $x \equiv p_{\rm tr}/\epsilon_{\rm out}$ and $y \equiv\Delta\epsilon/\epsilon_{\rm out}$. }
\label{mpc}
\end{figure}

\begin{figure}
\includegraphics[width=0.45\textwidth]{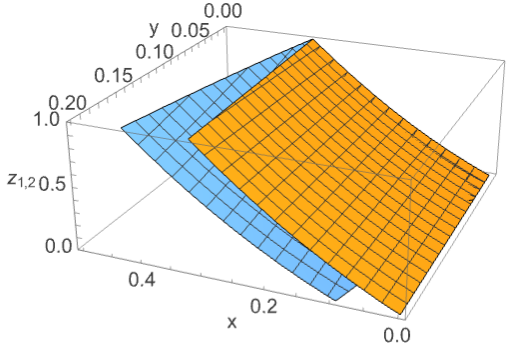}
\caption{Plot of the $z$ coordinate for each of the 2 threshold points $z_{1,2}$ (where $M'(z)=0$) as a function of the parameters $x$ and $y$. One can see they coincide only when $y\sim \Delta\epsilon =0$ or $x\sim p_{\rm tr} =0$.}
\label{deltaz}
\end{figure}

Before introducing the formalism itself, we point out its main characteristic: the coalescence of equilibrium solutions. Indeed, we are able to write the mass of the star as a function of the energy gap $\Delta \epsilon$, the transition pressure $p_{\rm tr}$ and the central pressure $p_c$. Written in terms of dimensionless quantities
\begin{equation}
x \equiv \frac{p_{\rm tr}}{\epsilon_{\rm out}} \quad
y \equiv \frac{\Delta\epsilon}{\epsilon_{\rm out}} \quad
z \equiv \frac{p_c}{\epsilon_{\rm out}} \quad
\tilde M \equiv \sqrt{\epsilon_{\rm out}}M 
\label{xyz}
\end{equation}
it yields
\begin{align}
\tilde M(z) = \frac{1}{4} \sqrt{\frac{3}{2\pi}} y 
\bigg[ 1-\frac{1-8\pi k}{(1+3x)^{4/3}} \bigg]^{1/4}+ \frac{4\pi}{3}k^3 \,,
\label{mtil}
\end{align}
where 
\begin{equation}
\frac{8\pi}{3}k^2 \equiv 
1-\bigg[\frac{(1+3x)(1+y+3x)}{1+y+3z}\bigg]^{4/3}
 \end{equation}
is a function of the previously defined variables $x$, $y$ and $z$. The plots $M \times p_c$ (actually, the corresponding dimensionless variables $\tilde M \times z$) are displayed in Fig.~\ref{mpc} for different values of the parameters $x$ and $y$.  Recall that the threshold points $z_j$ (where ${\tilde M}'=0$) splits the curve (i.e., the equilibrium configurations) into stable (${\tilde M}'>0$) and unstable (${\tilde M}'<0$) ones.  As the parameters $x$ and $y$ change, the curve itself (i.e., the equilibrium configurations) does change but, more importantly, the relative position $\Delta z\equiv |z_2-z_1|$ of the threshold points does change as well.  From Eq.~(\ref{mtil}), one is able to calculate it exactly, but it suffices to write its first-order term in a series expansion for small $x$, $y$ (i.e, small transition pressure with a small energy gap):
\begin{equation}
\Delta z = \sqrt{6 \pi} \sqrt{x} y
\end{equation}
to show the behavior of the threshold points $z_{1,2}$ in Fig.~\ref{deltaz}.

One can then see that they coalesce only for $x\sim p_{\rm tr}=0$ and/or $y\sim \Delta \epsilon=0$. Recalling their definitions above --- see Eq.~(\ref{xyz}) --- such solutions correspond to a homogeneous star (with either $\epsilon=\epsilon_{\rm in}$ or $\epsilon=\epsilon_{\rm out}$). Indeed, such a trivial example does correspond to trivial solutions. In the next Section we will see how one deals with such phenomenon in more general terms.

\section{Basic elements of Catastrophe Theory}
\label{sec:catastrophe}

Catastrophe Theory deals with the  changes in the number of equilibrium solution in a given system due to a (small) smooth variation of the so-called control parameters. For the sake of the argument, we will assume the configuration of the system is set by a single variable $z$ and that there are only 2 control parameters $\{\alpha,\beta\}$\footnote{Generalizations for higher number of both objects can be seen in Ref.~\cite{saunders1980catastrophe}}.

\begin{figure}
\centering
\includegraphics[width=0.4\textwidth]{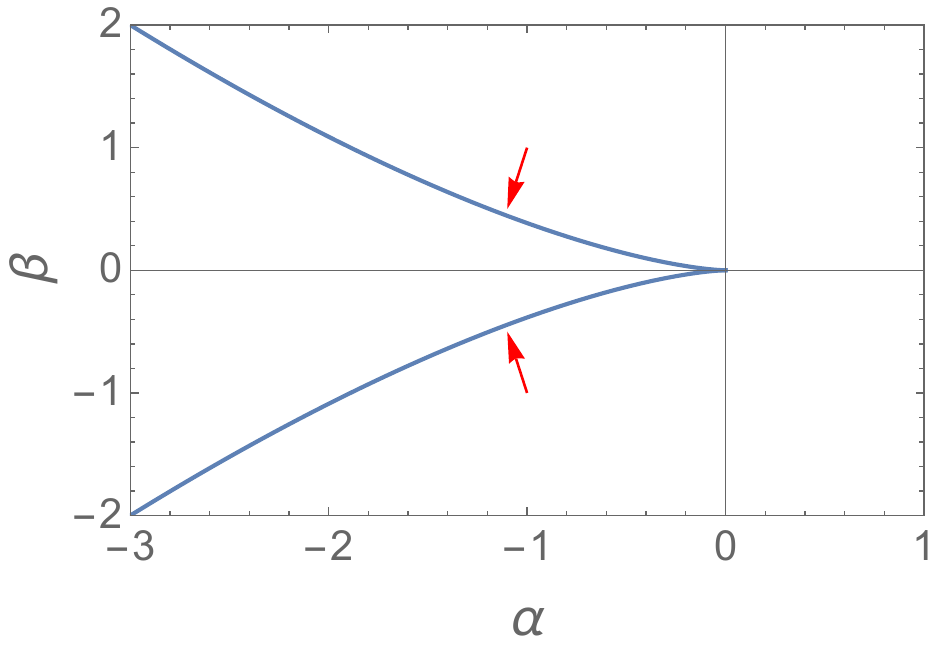}
\begin{picture}(0,0)
\put(-60,80){\includegraphics[width=0.12\textwidth]{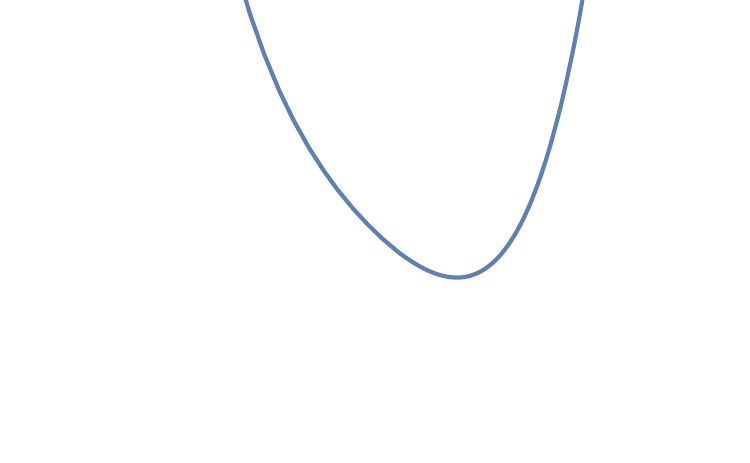}}
\put(-55,30){\includegraphics[width=0.12\textwidth]{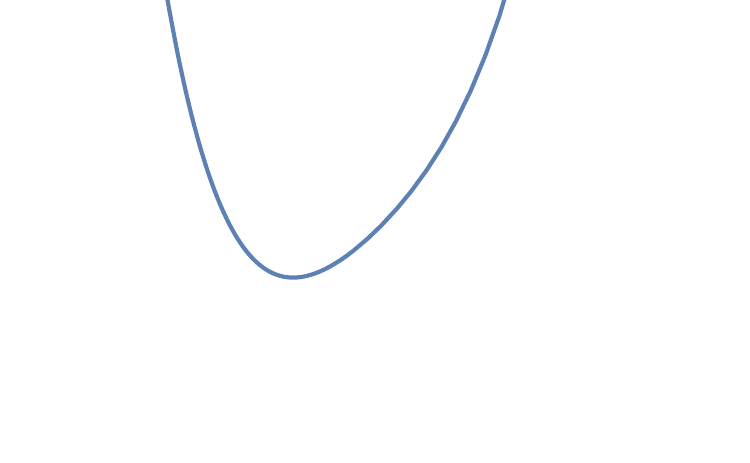}}
\put(-180,85){\includegraphics[width=0.08\textwidth]{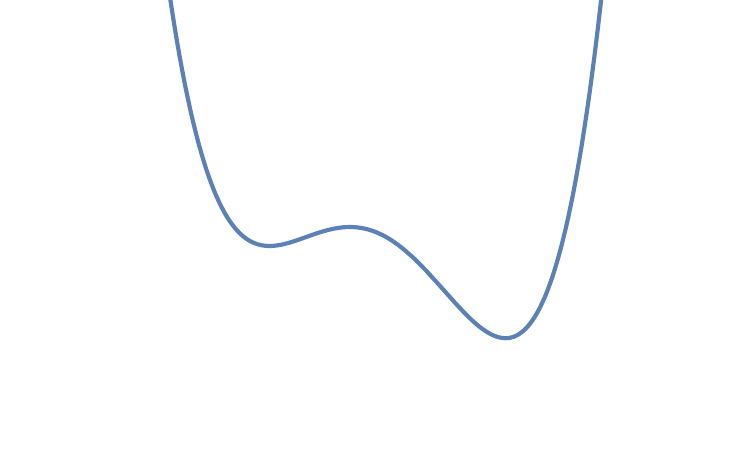}}
\put(-180,40){\includegraphics[width=0.08\textwidth]{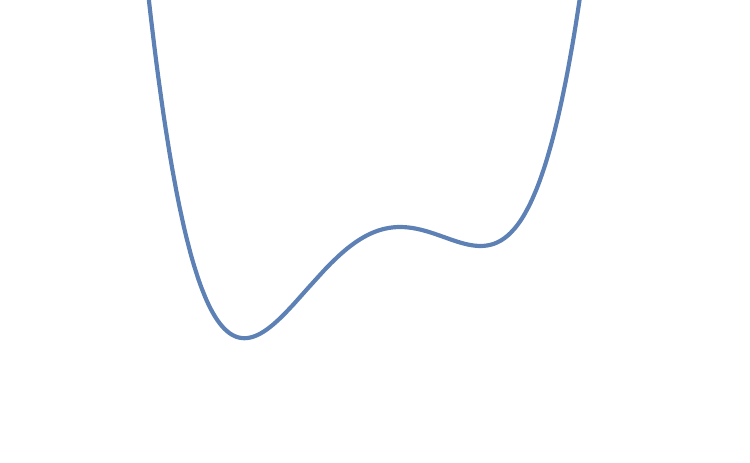}}
\put(-95,100){\includegraphics[width=0.06\textwidth]{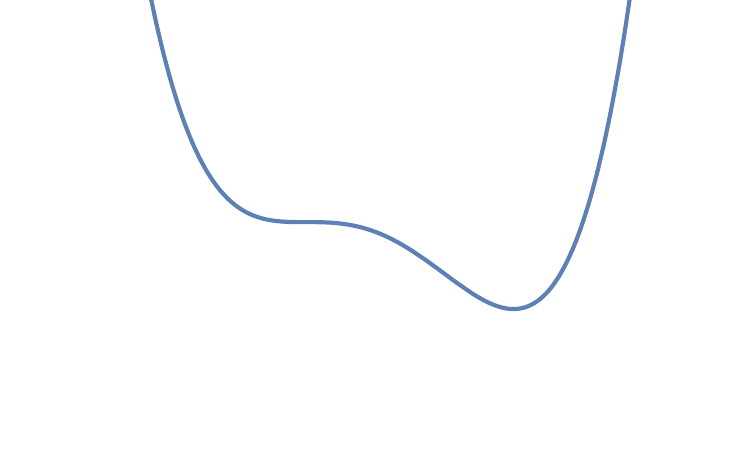}}
\put(-90,40){\includegraphics[width=0.06\textwidth]{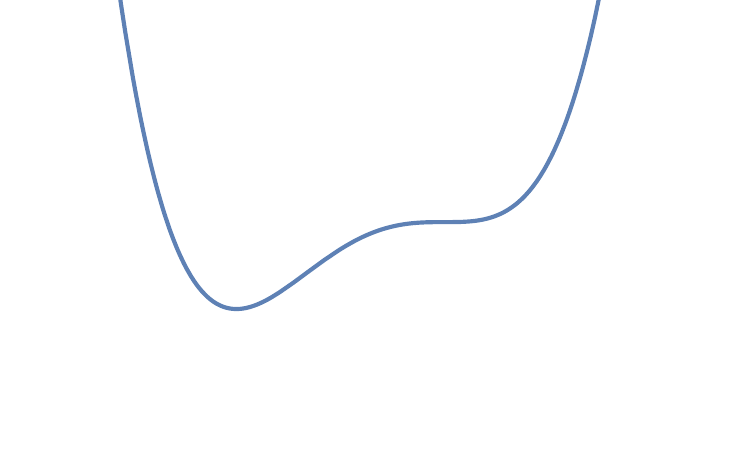}}
\end{picture}
\caption{Plot of the fold given by Eq.~(\ref{fold}) and the corresponding form of the potential energy, Eq.~(\ref{Vx}),  in each region. The middle column of insets indicates the form of the potential {\it on} the bifurcation set, as indicated by the arrows. Inside the set, the potential has 3 extrema; outside, only one.}
\label{foldplot}
\end{figure}

The equilibrium solutions can be seen as the ones that extremize a suitable potential energy $V(z)$. A simple  and general expression\footnote{One can get rid of a possible cubic term by a suitable shift on $z$.} that does the job is
\begin{equation}
V(z; \alpha,\beta) \equiv 
\frac{1}{4} z^4 + \frac{\alpha}{2} z^2 + \beta z \, ,
\label{Vx}
\end{equation}
where we have made explicit the dependence on the control parameters. 
In the region of the parameter space where $V$ features three extrema, there are three real solutions to $dV/dz=0$. If the potential energy is bounded from below, it is easy to see that one of them is necessarily a local maximum (unstable equilibrium), another one is a local minimum (stable equilibrium) and the last one is the global minimum (also stable equilibrium). Outside that region, there is only one extremum: a minimum.
The boundary between those two regions is defined, then, as the curve in the parameter space where the local maximum and the local minimum coalesce. Mathematically, it is defined by the so-called {\it singularity set}:
\begin{equation}
\frac{d^2V}{dz^2}\bigg|_{z_{\rm eq}} = 0 \, ,
\label{singset}
\end{equation}
where $z_{\rm eq}$ is one of the extrema, given, of course, by 
\begin{equation}
\frac{dV}{dz}\bigg|_{z_{\rm eq}} = 0,
\label{eqset}
\end{equation}
which is the so-called {\it equilibrium set}. The solution of Eqs.~(\ref{singset}) and (\ref{eqset}) is the so-called {\it bifurcation set}, given by, in this example:
\begin{equation}
4 \alpha^3 + 27 \beta^2 =0 \, .
\label{fold}
\end{equation}
The potential energy given by Eq.~(\ref{Vx}) has only one Real extremum (a minimum) if the parameters lie outside the curve given by Eq.~(\ref{fold}) in the parameter space $\{\alpha,\beta\}$. Inside it, there are 3 extrema, 2 of which are minima and 1 a maximum.  On the curve itself\footnote{ The roots do not annihilate each other upon crossing the bifurcation set --- they actually become complex conjugated, but we will ignore them since complex configurations do not correspond to physical systems.}, two of them coalesce --- which ones is set by the sign of $\beta$. Fig.~\ref{foldplot} shows the bifurcation set in the control-parameter space and the corresponding sketches of the potential $V(z; \alpha,\beta)$ in each region. 
We should point out that the values of $z_{\rm eq}$ do change with the variation of the control parameters, even without crossing the singularity set. Nevertheless, the {\it structural form}, i.e., the number of extrema, does remain the same in each region. This is clearly exemplified in the case of the semiclassical partition function for the double-well potential \cite{deCarvalho:2001vk,Kroff:2013jya}.

\begin{figure}[t]
\includegraphics[width=0.4\textwidth]{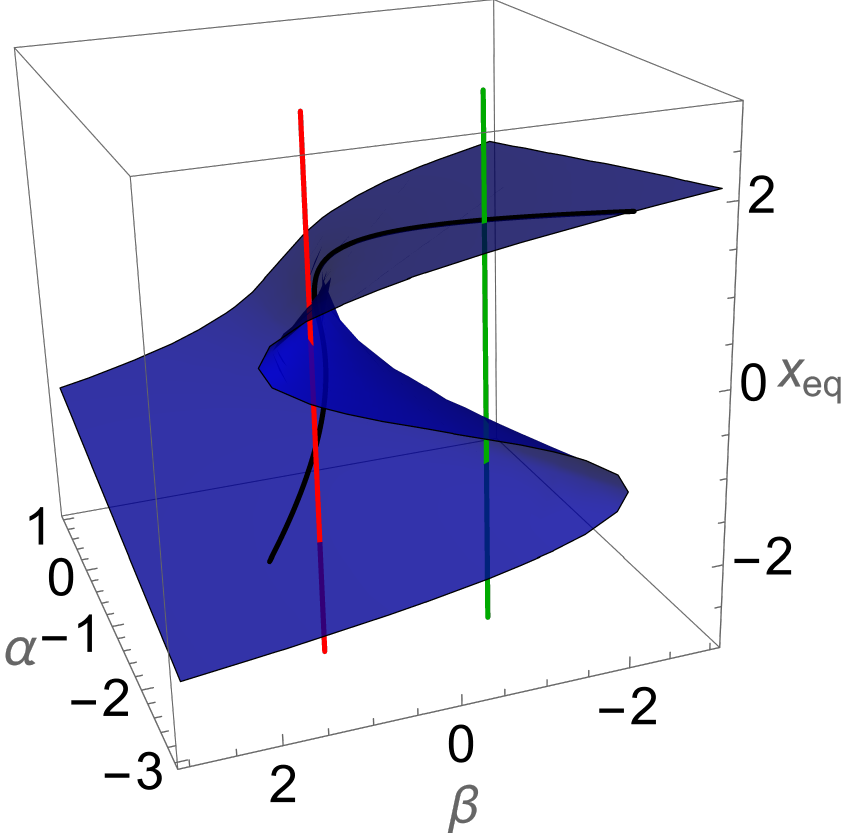}
\caption{Plot of the equilibrium positions, given by Eq.~(\ref{eqset}) as a function of the control parameters $\{\alpha,\beta\}$. The black solid curve is the projection of the fold (\ref{fold}) onto that surface. The vertical red and green straight lines are arbitrary, except for both being inside the fold. Each one crosses three times the equilibrium surface for each pair $\{\alpha,\beta\}$ (if inside the fold). By moving the red (green) line to the left (right), one can picture which solutions coalesce and which one survives when the fold is crossed.}
\label{fold3D_lines}
\end{figure}

In Fig.~\ref{fold3D_lines}, one can see the equilibrium surface (i.e., the values of the equilibrium points $z_{\rm eq}$) as a function of the control parameters $\{\alpha,\beta\}$. The fold is projected onto this surface as a black curve. One can see that, inside the fold, vertical lines (fixed values of the control parameters) cross the surface three times, each for a given equilibrium solution. If we shift the red (green) line towards larger (smaller) values of $\beta$, i.e., towards the closest fold branch, then the largest (smallest) values of $x_{eq}$ coalesce.  

\begin{figure}[t]
\includegraphics[width=0.4\textwidth]{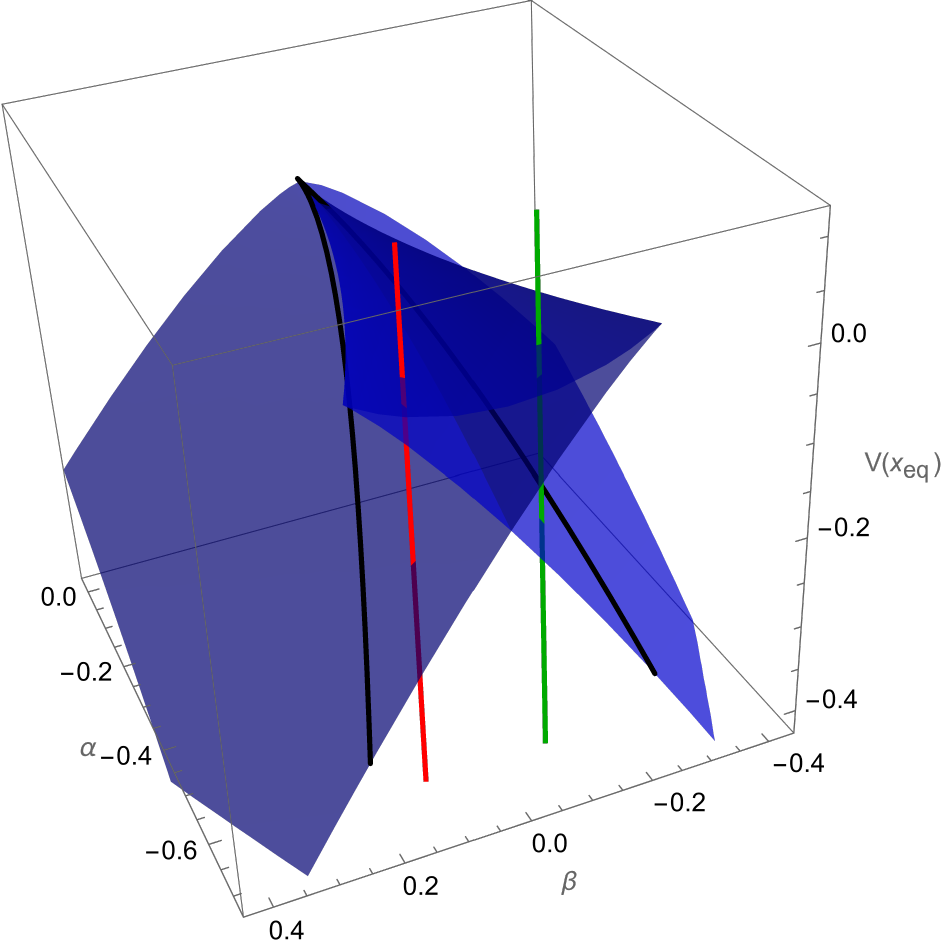}
\caption{Plot of the potential (\ref{Vx}),  calculated at the equilibrium positions $z_{eq}$ given by Eq.~(\ref{eqset}), as a function of the control parameters $\{\alpha,\beta\}$. The black solid curve is the projection of the fold (\ref{fold}) onto that surface. Vertical red and green lines as in the previous figure.}
\label{swallow3D_lines}
\end{figure}

Accordingly, we can plot the so-called swallowtail curve also as a function of $\{\alpha,\beta\}$ --- see Fig.~\ref{swallow3D_lines}. One can see that the highest $V(z_{eq})$ is always unstable, as expected ($d^2V(z_{\rm eq})/dz^2<0$). Here, as in the previous plot, when we shift the red (green) vertical line towards the nearest fold branch, it is always the equilibrium point corresponding to the highest $V(z_{eq})$, the local maximum, that coalesces with the one corresponding to the intermediate value $V(z_{eq})$, i.e., the local minimum. Outside the fold, only the global minimum survives.

\section{Connecting the dots: catastrophe theory for the stability of  stars}
\label{sec:dots}

\subsection{Standard relativistic stars }

The crucial first step is the definition of the function $V(z;\alpha,\beta)$ for the problem at hand. 

The standard mass $M$ {\it versus} central energy density $\epsilon_c$ plot of a relativistic non-hybrid  star in GR is a convex function near the tipping point, $\epsilon_c=\epsilon_{c*}$, that separates the stable equilibrium configurations (lower central energy densities, where $dM/d\epsilon_c>0$) from unstable (higher ones, where $dM/d\epsilon_c<0$). Let us focus on configurations around $\epsilon_{c*}$, where $M(\epsilon_{c*})=M_{\rm max}$, and approximate $M(\epsilon_c)$ by a quadratic function. 

In that region, there are two solutions for a given $M$: one stable and one unstable. The ``potential energy" $V(z;\alpha,\beta)$ should then feature two extrema and can therefore be written as a cubic function of $z\equiv \epsilon_c$. Notice that, as in the previous section\footnote{Although the function $V(z)$ is actually quartic --- see, in particular, Fig.~\ref{foldplot}.}, the values of $z_{\rm eq}$, i.e., the equilibrium configurations,  do change with the control parameters (along the (un)stable branches), as it should, but the structural from of the potential $V(z;\alpha,\beta)$ remains the same. In this simple case, we can associate the mass as a function of the central density, $M(\epsilon_c)$, to the {\it derivative} of the potential function $V(z;\alpha)$ with a single control parameter  $\alpha$, which corresponds to the {\it fold} catastrophe. The mass itself is related to the control parameter, since it dictates the existence (or lack thereof) of two ($M<M_{\rm max}$) or no equilibrium solutions (if $M>M_{\rm max}$), and the coalescence of both (at $M=M_{\rm max}$).


\subsection{Hybrid stars}


In a more realistic description of hybrid stars, one cannot assume two homogeneous layers. One has to pick a given equation of state for each layer --- such as two polytropic expressions, for instance:
\begin{align}
p(\epsilon) =& \Theta(p-p_{\rm tr}) \left(\Delta \epsilon +\left(\frac{p}{\kappa }\right)^{1/\gamma }\right)+\\
&+\Theta (p_{\rm tr}-p) \left(\frac{p}{\kappa }\right)^{1/\gamma },
\end{align}
where $\gamma=2$ and $\kappa=10^8~m^2$.
The particular choice itself is not important for the coalescence of threshold points, which relies only on the parameters already pointed out: $p_{\rm tr}$ and $\Delta \epsilon$, i.e., our $x$ and $y$ dimensionless variables --- all of which are present in the above equation.

\begin{figure}[t]
\includegraphics[width=0.45\textwidth]{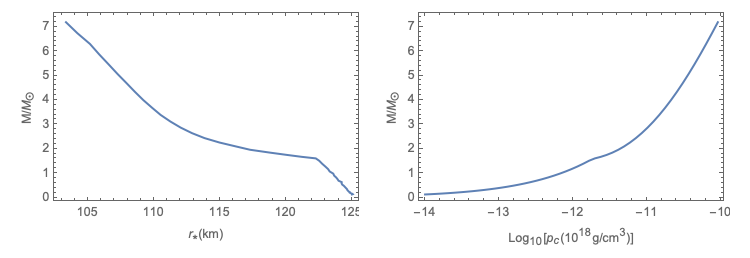}
\includegraphics[width=0.45\textwidth]{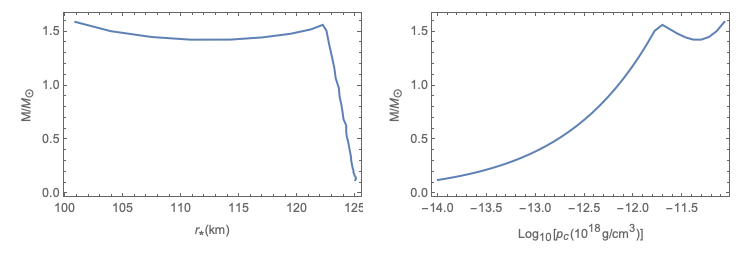}
\caption{Plots $M \times r_*$ {(left column)} and $M \times p_c$ {(right column)} for of $p_{\rm tr}/\epsilon_{\rm out} = 10^{-4}$ and $\Delta\epsilon/\epsilon_{\rm out} = 0.2$ {(top row)} or $1$ {(bottom row)}. One can see that the unstable solution disappears in the top row.}
\label{cusp}
\end{figure}

Nevertheless, the structure of the mass-central density diagram sketched in Ref.~\cite{Alford:2013aca} points out to a richer phenomenon.  In Fig.~\ref{cusp} one can see that the unstable branch disappears in the top row. On the other hand, in the bottom row, there are no solutions above the local maximum mass --- which corresponds to the coalescence of the stable and unstable branches.

First, one must notice that the hadronic branch (solid green line in that reference) is always present (since the pressure will eventually become smaller than $p_{\rm tr}$ before reaching the surface, where $p=0$) and, therefore, it does not play a role in the forthcoming discussion about the coalescence of solutions. 

Discarding this branch throughout the diagram, one can identify 4 equilibrium (either stable or unstable) solutions in the central region of the diagram. Hence, the associated potential energy $V(z)$ introduced in the previous Section must have 4 extrema (being 2 local maxima and, therefore, unstable); it is, thus, a $z^5$ polynomial. The corresponding normal form is the so-called swallowtail, given by
\begin{equation}
V(z; \alpha,\beta,\gamma) = \frac{1}{5} z^5 + \frac{1}{3} \alpha z^3 +  \beta z^2 + \gamma  z  \, , 
\end{equation}
with 3 control parameters ($\alpha$, $\beta$ and $\gamma$) --- two of which are related to ours $x$ and $y$ (respectively, normalized $p_{\rm tr}$ and $\Delta \epsilon$).

The equilibrium set is given by the $4$ roots of
\begin{equation}
 0 = V'(z) = z^4 +  \alpha z^3 +  \beta z + \gamma  \, .
\end{equation}

\begin{figure}[ht]
\includegraphics[width=0.4\textwidth]{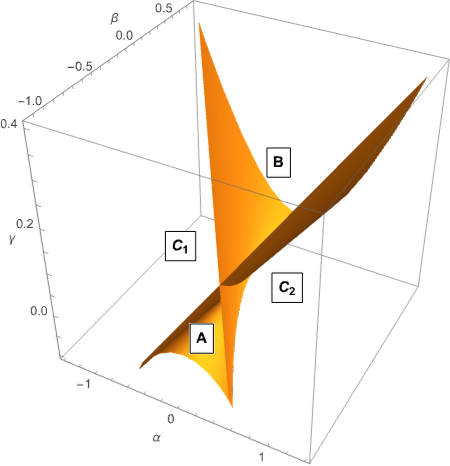}
\caption{Bifurcation set for the swallowtail catastrophe.}
\label{swallowtail}
\end{figure}

It is not possible to draw the equilibrium surface (i.e., the set of points $z_{\rm eq}$ where $dV/dz=0$) in terms of the 3 parameters  ($\alpha,\beta,\gamma$), since it is a 4D surface. We can, however, plot the bifurcation set (the loci of points in the parameter space where $dV/dz=0=d^2V/dz^2$) --- see Fig.~\ref{swallowtail}. The number of extrema changes as this surface is crossed. In the central limited region ($A$) there are 4 extrema: 2 maxima and 2 minima. They correspond, respectively, to the 2 dashed red lines (unstable hybrid stars) and to the 2 full red lines (stable hybrid stars) in Ref. \cite {Alford:2013aca}.
In region $B$, there are no extrema (except for the hadronic solution, which we are neglecting from the beginning). In regions $C_1$ and $C_2$ there is only 1 pair of extrema (1 maximum, 1 minimum), since the other pair has merged. Notice that which pair disappears depends on which region is reached (hence the index in $C_{1,2}$). 

The positions of the extrema --- that characterize the physical quantities of a given star (central pressure, radius, mass) --- do change if we move around inside each of those regions, but the number of such points does not.

\begin{figure}[ht]
\includegraphics[width=0.4\textwidth]{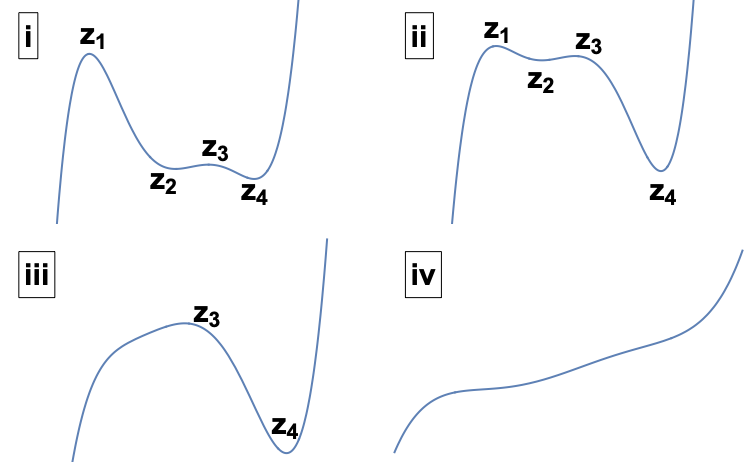}
\caption{Sketch of the potential $V(z)$ for different regions of the bifurcation set. Panels {\bf (i), (ii)} correspond to Region A; note the presence of 2 local maxima ($z_1$ and $z_3$) and 2 minima ($z_2$ and $z_4$). From panel {\bf (ii)} to panel {\bf (iii)}, the leftmost maximum ($z_1$) and the leftmost minimum ($z_2$) have merged when the surface between $A$ and $C_1$ is crossed. In panel {\bf (iv)}, Region B, all the extrema have coalesced; there is no extreme point.}
\label{vabcd}
\end{figure}

If we label the threshold points in the central region by $z_1, \cdots , z_4$ as illustrated in Fig. \ref{vabcd}, we can see that a pair of neighboring extrema coalesce whenever the control parameters leave Region $A$. 

\section{Summary and outlook}
\label{sec:summary}

We have discussed the possible topologies of the mass-radius relation for hybrid stars in light of catastrophe theory, assuming a single first-order phase transition between nuclear and quark matter. As a general feature, we showed that different branches of compact stars emerge or disappear depending on the bifurcation structure of the underlying catastrophe potential function, the mathematical analog of an effective potential. 

After discussing the simpler case of incompressible fluids, which we call bi-homogeneous hybrid stars, where most of the calculations can be performed analytically, we presented a more general, qualitative discussion explaining the (dis)appearance of (un)stable branches for hybrid stars as the control parameters change. Considering the topology of the mass-central energy density diagram sketched in Ref. \cite{Alford:2013aca}, we built a qualitative description of the stability structure based on the normal form of the swallowtail catastrophe.

In the near future, we shall report on the analysis, via catastrophe theory, of the stability of hybrid stars generated by realistic equations of state. Then, quantitative predictions regarding the different sectors of the bifurcation set will be presented. In particular, we will be able to tell if the regions $C_1$ and $C_2$ yield stars with different characteristics or if they are totally equivalent.

\begin{acknowledgments}
This work was partially supported by INCT-FNA (Process No. 464898/2014-5), CAPES (Finance Code 001), CNPq, and FAPERJ.
\end{acknowledgments}

\bibliography{references}

@article{deCarvalho:2001vk,
    author = "de Carvalho, C. A. A. and Cavalcanti, R. M. and Fraga, E. S. and Joras, S. E.",
    title = "{Improved semiclassical density matrix: Taming caustics}",
    eprint = "quant-ph/0105076",
    archivePrefix = "arXiv",
    doi = "10.1103/PhysRevE.65.056112",
    journal = "Phys. Rev. E",
    volume = "65",
    pages = "056112",
    year = "2002"
}

@article{Kroff:2013jya,
    author = "Kroff, D. and Bessa, A. and de Carvalho, C. A. A. and Fraga, E. S. and Jor{\'a}s, S. E.",
    title = "{Semiclassical partition function for the double-well potential}",
    eprint = "1311.1182",
    archivePrefix = "arXiv",
    primaryClass = "hep-ph",
    doi = "10.1103/PhysRevD.90.025019",
    journal = "Phys. Rev. D",
    volume = "90",
    number = "2",
    pages = "025019",
    year = "2014"
}

@article{Ramsey1950,
  author  = {Ramsey, W. H.},
  title   = {On the Instability of Small Planetary Cores (I)},
  journal = {Monthly Notices of the Royal Astronomical Society},
  volume  = {110},
  pages   = {325--338},
  year    = {1950},
  doi     = {10.1093/mnras/110.4.325},
  adsurl  = {https://ui.adsabs.harvard.edu/abs/1950MNRAS.110..325R}
}

@article{Lighthill1950,
  author  = {Lighthill, M. J.},
  title   = {On the Instability of Small Planetary Cores (II)},
  journal = {Monthly Notices of the Royal Astronomical Society},
  volume  = {110},
  number  = {4},
  pages   = {339--342},
  year    = {1950},
  month   = {Aug},
  doi     = {10.1093/mnras/110.4.339},
  adsurl  = {https://ui.adsabs.harvard.edu/abs/1950MNRAS.110..339L}
}

@article{deforcrand2009pos,
  author        = {de Forcrand, Philippe},
  title         = {Simulating QCD at finite density},
  journal       = {Proceedings of Science (PoS)},
  volume        = {LAT2009},
  pages         = {010},
  year          = {2009},
  doi           = {10.22323/1.091.0010},
  eprint        = {0905.4268},
  archivePrefix = {arXiv},
  primaryClass  = {hep-lat}
}

@article{Hands:2007by,
    author = "Hands, Simon",
    editor = "Kunihiro, Teiji and others",
    title = "{Simulating dense matter}",
    eprint = "hep-lat/0703017",
    archivePrefix = "arXiv",
    doi = "10.1143/PTPS.168.253",
    journal = "Prog. Theor. Phys. Suppl.",
    volume = "168",
    pages = "253--260",
    year = "2007"
}

@book{saunders1980catastrophe,
  author    = {Saunders, P. T.},
  title     = {An Introduction to Catastrophe Theory},
  publisher = {Cambridge University Press},
  year      = {1980},
  address   = {Cambridge}
}

@article{Kampfer:1981zmq,
    author = {K{\"a}mpfer, B.},
    title = "{On stabilizing effects of relativity in cold spheric stars with a phase transition in the interior}",
    doi = "10.1016/0370-2693(81)90065-4",
    journal = "Phys. Lett. B",
    volume = "101",
    pages = "366--368",
    year = "1981"
}

@article{Seidov1967,
  author    = {Seidov, Z. F.},
  title     = {The Equilibrium of a Star during a Phase Transition},
  journal   = {Astrophysics},
  volume    = {3},
  number    = {2},
  pages     = {90--95},
  year      = {1967},
  month     = {Jun},
  doi       = {10.1007/BF01013200}
}

@article{seidov1971stability,
  author  = {Seidov, Z. F.},
  title   = {The Stability of a Star with a Phase Change in General Relativity Theory},
  journal = {Soviet Astronomy},
  volume  = {15},
  pages   = {347},
  year    = {1971},
  month   = oct
}

@article{schaeffer1983phase,
  author  = {Schaeffer, R. and Zdunik, L. and Haensel, P.},
  title   = {Phase transitions in stellar cores. I - Equilibrium configurations},
  journal = {Astronomy and Astrophysics},
  volume  = {126},
  pages   = {121--145},
  year    = {1983},
  month   = sep
}

@article{Zdunik1987,
  author    = {Zdunik, J. L. and Haensel, P. and Schaeffer, R.},
  title     = {Phase transitions in stellar cores. II. Equilibrium configurations in general relativity},
  journal   = {Astronomy and Astrophysics},
  volume    = {172},
  number    = {1-2},
  pages     = {95--110},
  year      = {1987},
  month     = {Jan}
}

@article{lindblom1998phase,
  author  = {Lindblom, Lee},
  title   = {Phase transitions and the mass radius curves of relativistic stars},
  journal = {Physical Review D},
  volume  = {58},
  pages   = {024008},
  year    = {1998},
  eprint  = {gr-qc/9802072},
  archivePrefix = {arXiv}
}

@book{harrison1965gravitation,
  title     = {Gravitation Theory and Gravitational Collapse},
  author    = {Harrison, B. Kent and Thorne, Kip S. and Wakano, Masami and Wheeler, John Archibald},
  year      = {1965},
  publisher = {University of Chicago Press},
  address   = {Chicago}
}

@book{Weinberg:1972kfs,
    author = "Weinberg, Steven",
    title = "{Gravitation and Cosmology}: {Principles and Applications of the General Theory of Relativity}",
    isbn = "978-0-471-92567-5, 978-0-471-92567-5",
    publisher = "John Wiley and Sons",
    address = "New York",
    year = "1972"
}

@article{Fraga:2018cvr,
    author = "Fraga, Eduardo S. and Hippert, Maur{\'\i}cio and Schmitt, Andreas",
    title = "{Surface tension of dense matter at the chiral phase transition}",
    eprint = "1810.13226",
    archivePrefix = "arXiv",
    primaryClass = "hep-ph",
    doi = "10.1103/PhysRevD.99.014046",
    journal = "Phys. Rev. D",
    volume = "99",
    number = "1",
    pages = "014046",
    year = "2019"
}

@article{Lugones:2013ema,
    author = "Lugones, G. and Grunfeld, A. G. and Al Ajmi, M.",
    title = "{Surface tension and curvature energy of quark matter in the Nambu-Jona-Lasinio model}",
    eprint = "1308.1452",
    archivePrefix = "arXiv",
    primaryClass = "hep-ph",
    doi = "10.1103/PhysRevC.88.045803",
    journal = "Phys. Rev. C",
    volume = "88",
    number = "4",
    pages = "045803",
    year = "2013"
}

@article{Mintz:2012mz,
    author = "Mintz, Bruno W. and Stiele, Rainer and Ramos, Rudnei O. and Schaffner-Bielich, Juergen",
    title = "{Phase diagram and surface tension in the three-flavor Polyakov-quark-meson model}",
    eprint = "1212.1184",
    archivePrefix = "arXiv",
    primaryClass = "hep-ph",
    doi = "10.1103/PhysRevD.87.036004",
    journal = "Phys. Rev. D",
    volume = "87",
    number = "3",
    pages = "036004",
    year = "2013"
}

@article{Pinto:2012aq,
    author = "Pinto, Marcus B. and Koch, Volker and Randrup, Jorgen",
    title = "{The Surface Tension of Quark Matter in a Geometrical Approach}",
    eprint = "1207.5186",
    archivePrefix = "arXiv",
    primaryClass = "hep-ph",
    doi = "10.1103/PhysRevC.86.025203",
    journal = "Phys. Rev. C",
    volume = "86",
    pages = "025203",
    year = "2012"
}

@article{Palhares:2010be,
    author = "Palhares, Leticia F. and Fraga, Eduardo S.",
    title = "{Droplets in the cold and dense linear sigma model with quarks}",
    eprint = "1006.2357",
    archivePrefix = "arXiv",
    primaryClass = "hep-ph",
    doi = "10.1103/PhysRevD.82.125018",
    journal = "Phys. Rev. D",
    volume = "82",
    pages = "125018",
    year = "2010"
}

@article{Schmitt:2020tac,
    author = "Schmitt, Andreas",
    title = "{Chiral pasta: Mixed phases at the chiral phase transition}",
    eprint = "2002.01451",
    archivePrefix = "arXiv",
    primaryClass = "hep-ph",
    doi = "10.1103/PhysRevD.101.074007",
    journal = "Phys. Rev. D",
    volume = "101",
    number = "7",
    pages = "074007",
    year = "2020"
}

@article{Fraga:2023wtd,
    author = {Fraga, Eduardo S. and da Mata, Rodrigo and Schaffner-Bielich, J{\"u}rgen},
    title = "{SU(3) parity doubling in cold neutron star matter}",
    eprint = "2309.02368",
    archivePrefix = "arXiv",
    primaryClass = "hep-ph",
    doi = "10.1103/PhysRevD.108.116003",
    journal = "Phys. Rev. D",
    volume = "108",
    number = "11",
    pages = "116003",
    year = "2023"
}

@article{Christian:2017jni,
    author = {Christian, Jan-Erik and Zacchi, Andreas and Schaffner-Bielich, J{\"u}rgen},
    title = "{Classifications of Twin Star Solutions for a Constant Speed of Sound Parameterized Equation of State}",
    eprint = "1707.07524",
    archivePrefix = "arXiv",
    primaryClass = "astro-ph.HE",
    doi = "10.1140/epja/i2018-12472-y",
    journal = "Eur. Phys. J. A",
    volume = "54",
    number = "2",
    pages = "28",
    year = "2018"
}

@article{Alford:2013aca,
    author = "Alford, Mark G. and Han, Sophia and Prakash, Madappa",
    title = "{Generic conditions for stable hybrid stars}",
    eprint = "1302.4732",
    archivePrefix = "arXiv",
    primaryClass = "astro-ph.SR",
    doi = "10.1103/PhysRevD.88.083013",
    journal = "Phys. Rev. D",
    volume = "88",
    number = "8",
    pages = "083013",
    year = "2013"
}

@book{glendenning1997compact,
  title     = {Compact Stars: Nuclear Physics, Particle Physics, and General Relativity},
  author    = {Glendenning, Norman K.},
  year      = {1997},
  edition   = {1},
  publisher = {Springer},
  series    = {Astronomy and Astrophysics Library},
  address   = {New York},
  isbn      = {978-0-387-94957-4}
}

@book{Schaffner-Bielich:2020psc,
    author = {Schaffner-Bielich, J{\"u}rgen},
    title = "{Compact Star Physics}",
    doi = "10.1017/9781316848357",
    isbn = "978-1-316-84835-7, 978-1-107-18089-5",
    publisher = "Cambridge University Press",
    month = "8",
    year = "2020"
}

@article{Tolman:1939jz,
    author = "Tolman, Richard C.",
    title = "{Static solutions of Einstein's field equations for spheres of fluid}",
    doi = "10.1103/PhysRev.55.364",
    journal = "Phys. Rev.",
    volume = "55",
    pages = "364--373",
    year = "1939"
}

@article{Oppenheimer:1939ne,
    author = "Oppenheimer, J. R. and Volkoff, G. M.",
    title = "{On massive neutron cores}",
    doi = "10.1103/PhysRev.55.374",
    journal = "Phys. Rev.",
    volume = "55",
    pages = "374--381",
    year = "1939"
}

\end{document}